\documentclass[preprintnumbers,amsmath,aps,nofootinbib,amssymb,superscriptaddress]{revtex4}
\usepackage{amsmath,amssymb}
\usepackage{graphicx}% Include figure files
\usepackage[english]{babel}
\usepackage{bm}% bold math
\usepackage{enumerate}

\begin{document}

\title{Classical and quantum exact solutions for the anisotropic Bianchi type I in multi-scalar field cosmology with an exponential potential driven inflation}
\author{J. Socorro}
\email{socorro@fisica.ugto.mx}
\author{Omar E. N\'u\~nez}
\email{neophy@fisica.ugto.mx}
\affiliation{Departamento de  F\'{\i}sica, DCeI, Universidad de Guanajuato-Campus Le\'on, C.P. 37150, Le\'on, Guanajuato, M\'exico}
\author{Rafael Hern\'{a}ndez-Jim\'{e}nez}
\email{s1367850@sms.ed.ac.uk}
\affiliation{School of Physics and Astronomy, University of Edinburgh, Edinburgh, EH9 3FD, United Kingdom}

\begin{abstract}

 The anisotropic Bianchi type I in multi-scalar field cosmology is studied with a particular potential of the form $\rm V=  V_0 e^{-\left[\lambda_1 \phi_1 + \cdots + \lambda_n \phi_n \right]}\,,$ which emerges as a condition between the time derivatives of their corresponding momenta. Using the Hamiltonian formalism for the inflation epoch with a quintessence framework we find the exact solutions for the Einstein-Klein-Gordon (EKG) system with different scenarios specified by the parameter $\rm \lambda^2= \sum_{i=1}^n \lambda_i^2$. For the quantum scheme of this model, the corresponding Wheeler-DeWitt (WDW) equation is solved by applying an appropriate change of variables and suitable ansatz.

\end{abstract}

%\pacs{4.20.Fy, 4.20.Jb, 98.80.-k, 98.80.Hw}
\maketitle


\section{Introduction}

The inflation paradigm is considered the most accepted mechanism to explain many of the fundamental problems of the early stages in the evolution of our universe \cite{guth1981, linde1982, turner1981, starobinsky1980}, such as the flatness, homogeneity and isotropy observed in the present universe. Another important aspect of inflation is its ability to correlate cosmological scales that would otherwise be disconnected. Fluctuations generated during this early phase of inflation yield a primordial spectrum of density perturbation \cite{Starobinsky:1979ty,Mukhanov:1981xt,kodama, bassett}, which is nearly scale invariant, adiabatic and Gaussian, which is in agreement with cosmological observations \cite{Planck}. 

The single-field scalar models have been broadly used to describe the primordial expansion, the most phenomenological successful are those with a quintessence scalar field and slow-roll inflation \cite{barrow, andrew1998b, ferreira, copeland1, copeland2, copeland3, andrew2007, gomez, capone, kolb}. However, if another component is included, i.e. a multi-scalar field theory, it is also possible to produce an inflationary scenario  \cite{coley, Copeland:1999cs}, even if the fields are non interacting \cite{andrew2007}. Even more, the dynamical possibilities in  multi-field inflationary scenarios are considerably richer than those in single-field models, such as in the primordial inflation perturbations analysis or the assisted inflation as discussed in \cite{andrew1998a}, furthermore, the general assisted inflation as in \cite{Copeland:1999cs}. In this sense the multi-scalar fields cosmology is an attractive candidate to explain such phenomenon. 

Moreover, there are works where a more generalized multi-scalar model with $n$ scalar fields have been studied \cite{Yokoyama:2007dw,Miao Li:2007,Chiba:2008rp,Jinn-Ouk Gong:2018} and very few have exact solutions, such is the case of Genly Leon et al. \cite{genly}, where the authors introduce new dynamical degrees of freedom in a multi-scalar field cosmology in order to explain the observational phenomena.

Recent works have shown that multi-scalar field models are very fruitful when studying the early stages of the universe, such is the case in \cite{DeCross}, where the authors perform a semi-analytic study of preheating in inflationary models comprised of multiple scalar fields coupled nonminimally to gravity. In \cite{Hotinli:2017vhx} the authors show the sensitivity of the cosmological observables to the reheating phase following an inflation driven by many scalar fields, where they find that for certain decay rate, reheating following multi-field inflation can have a significant impact on the prediction of cosmological observables. 

The hypothesis of primordial anisotropy at early stages of the universe, and even predating inflation, is an enticing proposal that can shed some light in the anomalies found in the cosmic microwave background anisotropies on large angular scales, some serious attempts have been made \cite{Pereira:2007yy,Pitrou:2008gk,Pereira:2015pga,Gumrukcuoglu:2007bx}, where anisotropic cosmological models, mainly the Bianchi type I model, have been employed as a spacetime background in an early anisotropic but homogenus universe that experiences an isotropization at the onset of inflation, however, traces of such anisotropies would lead to the anomalies found in the thermal maps of the cosmic microwave background. In the post inflationary evolution the universe tends towards a Friedmann-Lema\^{i}tre-Robertson-Walker (FLRW) spacetime, therefore, the standard picture of the evolution of the universe is recovered. Indeed, anisotropic cosmological models, represent an alluring prospect to explain the early stages of the universe, even if no conclusive evidence has been found that a primordial anisotropy is needed.

There are many notable works on the subject of anisotropic cosmological model of inflation, such is the case in \cite{Folomeev:2007uw} where the author employs a Bianchi type I model with two interacting scalar fields and a potential energy of the form $\rm V(\phi, \chi)\sim \phi^4+\chi^4$, numerical solutions and the asymptotically isotropic Friedmann case are found. In previous works \cite{Socorro:2017dle, ssw} we have shown that an exponential potential of the form $\rm V(\phi, \sigma)\sim e^{\phi+\sigma}$ could be a viable candidate for inflation, in both flat isotropic and anisotropic spacetime. Even more cases of anisotropic cosmological models or exponential potential for inflation can be found in previous works \cite{gssa, socorro-doleire, socorro-pimentel}. 

In the present work we analyze the case of Bianchi type I model with multi-scalar field cosmology, constructed using all quintessence fields $\rm \{\phi_{1},\dots,\phi_{n}\}$, and an exponential potential $\rm V(\phi_1, \ldots ,\phi_n)$ in order to find exact solutions to Einstein-Klein-Gordon (EKG) equations. The Hamilton's formulation, which is widely used in analytical mechanics is employed, using this method we are able to obtain the exact solutions of the complete set of EKG equations without using any approximation.

On the other hand, we implement a basic formulation in quantum cosmology by means of the Wheeler-DeWitt (WDW) equation. The WDW equation has been analyzed with different approaches in order to solve it, and there are several papers on the subject, such is the case in \cite{Gibbons}, where the authors debate what a typical wave function for the universe is. In \cite{Zhi} there is a review on quantum cosmology where the problem of how the universe emerged from a big bang singularity can no longer be neglected in the GUT epoch. Moreover, the best candidates for quantum solutions are those that have a damping behavior with respect to the scale factor, since only such wave functions allow for good classical solutions when using a Wentzel-Kramers-Brillouin (WKB) approximation for any scenario in the evolution of our universe \cite{HH,H}.

This work is arranged as follows. In section \ref{model} we present the corresponding EKG equation for the multi-scalar fields model and the Einstein field equations for the anisotropic Bianchi type I cosmological model.
In section \ref{hap},  we introduce the hamiltonian formalism for this cosmological model in the representation for the radii as an exponential type, here the Lagrangian and Hamiltonian density are presented.
In section \ref{solutions} classic solutions are obtained for the dynamic equations from the Hamiltonian density, obtaining one relation between the momenta associated with the scalar fields $\rm \{\phi_{1},\dots,\phi_{n}\}$ which imply the structure for the scalar potential employed in this work, the exact solutions are found for different scenarios specified by the parameter $\rm \lambda^2= \sum_{i=1}^n \lambda_i^2$, the number of e-folds and the anisotropic density for all the cases are also presented. 
In section \ref{quantum} we present the WDW equation which is solved in a general way. 
%considering the factor ordering problem for the three radii, common in quantum cosmology. 
Finally, in section \ref{conclusions} we present our conclusions for this work.

\section{The model \label{model}}

We begin with the construction of a multi scalar fields cosmological paradigm, which requires n canonical scalar fields $\rm \{\phi_{1},\dots,\phi_{n}\}$. The action of a universe constituted of such fields is
\begin{equation}
\rm {\cal L}=\sqrt{-g} \left( R+\frac{1}{2} \sum_{i=1}^n g^{\mu\nu}\nabla_\mu \phi_i \nabla_\nu \phi_i-V(\phi_1, \dots ,\phi_n)\right) \,, \label{lagra}
\end{equation}
where $\rm R$ is the Ricci scalar, $\rm V(\phi_1, \ldots ,\phi_n)$ is the corresponding scalar field potential, and the reduced Planck mass $M_{P}^{2}=1/8\pi G=1$. The variations of eq.~(\ref{lagra}) with respect to the metric and the scalar fields give the Einstein-Klein-Gordon (EKG) field equations
\begin{eqnarray}
&& \rm G_{\alpha \beta} = \frac{1}{2}\left(\sum_{i=1}^n \nabla_\alpha \phi_i \nabla_\beta \phi_i -\frac{1}{2}\sum_{i=1}^n g_{\alpha \beta} g^{\mu \nu}\nabla_\mu \phi_i \nabla_\nu \phi_i \right)-\frac{1}{2}g_{\alpha \beta} \, V(\phi_1, \ldots ,\phi_n) \,,\label{munu}\\
&& \rm \Box \phi_i - \left( \frac{\partial V(\phi_1, \ldots ,\phi_n)}{\partial \phi_i} \right)_{\phi_{j \not= i}}=g^{\mu\nu} {\phi_i}_{,\mu\nu} -g^{\alpha \beta} \Gamma^\nu_{\alpha\beta} \nabla_\nu \phi_i - \left( \frac{\partial V}{\partial \phi_i} \right)_{\phi_{j \not= i}}=0 \,,\qquad i=1,\cdots,n \,.\label{ekg}
\end{eqnarray}
The line element to be considered in this work is the anisotropic Bianchi type I cosmological metric in the Misner's parameterization 
\begin{equation}\label{bianchi-i}
\rm ds^2=-N(t)^2 dt^2 +e^{2a(t)} dx^2 + e^{2b(t)} dy^2 +e^{2c(t)} dz^2 \,,
\end{equation}
where $\rm N(t)$ is the lapse function, which in a special gauge one can directly recover the cosmic time $\rm t_{phys}$ ($\rm Ndt=dt_{phys}$), and the scale factors $\rm A(t)=e^{a(t)}, B(t)=e^{b(t)}, C(t)=e^{c(t)} $ in the directions (x,y,z) respectively. Hence the EKG field equations are
\begin{eqnarray}
\rm \frac{\dot a \dot b}{N^2} +\frac{\dot a \dot c}{N^2} +\frac{\dot b \dot c}{N^2}-\frac{1}{4N^2}\sum_i^n\dot{\phi_i}^2-\frac{1}{2}V(\phi_1, \cdots \phi_n)&=&0 \,,
\label{ein0}\\
\rm \frac{\ddot b}{N^2}+\frac{\dot b^2}{N^2}+\frac{\ddot c}{N^2}+\frac{\dot c^2}{N^2}+\frac{\dot b \dot c}{N^2}  - \frac{\dot b \dot N}{N^3} -\frac{\dot c \dot N}{N^3} +\frac{1}{4N^2}\sum_i^n\dot{\phi_i}^2
-\frac{1}{2}V(\phi_1, \cdots \phi_n)&=&0 \,, \label{ein1}\\
\rm \frac{\ddot a}{N^2}+\frac{\dot a^2}{N^2}+\frac{\ddot c}{N^2}+\frac{\dot c^2}{N^2}+\frac{\dot a \dot c}{N^2} - \frac{\dot a \dot N}{N^3} -\frac{\dot c \dot N}{N^3} +\frac{1}{4N^2}\sum_i^n\dot{\phi_i}^2
-\frac{1}{2}V(\phi_1, \cdots \phi_n)&=&0 \,,\label{ein2}\\
\rm \frac{\ddot a}{N^2}+\frac{\dot a^2}{N^2}+\frac{\ddot b}{N^2}+\frac{\dot b^2}{N^2} +\frac{\dot a \dot b}{N^2}- \frac{\dot b \dot N}{N^3} -\frac{\dot a \dot N}{N^3} +\frac{1}{4N^2}\sum_i^n\dot{\phi_i}^2
-\frac{1}{2}V(\phi_1, \cdots \phi_n)&=&0 \,,  \label{ein3}\\
\rm -\frac{(\dot a +\dot b +\dot c)}{N}\frac{\dot \phi_i}{N} -\frac{ \ddot \phi_i}{N^2}+\frac{\dot N \dot \phi_i}{N^3}- \left( \frac{\partial V(\phi_1, \cdots \phi_n)}{\partial \phi_i}\right)_{\phi_{j \not= i}}&=& \rm 0, \quad i=1,\cdots , n \label{kg-phi}
\end{eqnarray}
where upper ``{\tiny{$\bullet$}}" represents the time derivatives. Due to the algebraic structure of above equations, one can take the difference between eqs.~(\ref{ein1}) and (\ref{ein2}), yielding the relation
\begin{equation}\label{first-relation}
\rm \ddot a -\ddot b + \dot a^2 -\dot b^2 + \dot c(\dot a -\dot b) - \frac{\dot N(\dot a - \dot b)}{N}=0 \,,
\end{equation}
then by defining $\rm {\cal U}=ABC=e^{a+b+c}$ we obtain as a solution of eq.~(\ref{first-relation}): 
\begin{equation}
\rm a=b+\kappa_1 \int \frac{Ndt}{{\cal U}}, \qquad \kappa_1=constant \,. \label{solution-a}
\end{equation}
Moreover, the other parameters $\rm (b,c)$ have similar relations
\begin{equation}\label{solution-b-c}
\rm b= c+\kappa_2 \int \frac{Ndt}{{\cal U}}\,,\quad c= a+\kappa_3 \int  \frac{Ndt}{{\cal U}}\,, \quad  \kappa_2=constant \,,\quad \kappa_3=constant
\end{equation}
%\begin{eqnarray}
%\rm b&=& \rm  c+\kappa_2 \int \frac{Ndt}{{\cal U}}, \qquad  \kappa_2=constant \,, %\label{solution-b}\\
%\rm c&=& \rm  a+\kappa_3 \int \frac{Ndt}{{\cal U}}, \qquad  \kappa_3=constant \,, %\label{solution-c}
%\end{eqnarray}
%
Notice that the constants satisfy $\rm \kappa_1+\kappa_2+\kappa_3=0$. Additionally, another combinations among the parameters (a,b,c), in term of the function $\rm {\cal U}$, are
\begin{equation}\label{solution2-a-b-c}
\rm a = ln \,{\cal U}^{1/3} -\frac{\kappa_2+2\kappa_3}{3} \int \frac{Ndt}{{\cal U}} \,,\quad b = \rm ln \,{\cal U}^{1/3} +\frac{\kappa_2-\kappa_1}{3} \int \frac{Ndt}{{\cal U}} \,,\quad c = \rm ln \,{\cal U}^{1/3} -\frac{\kappa_1+2\kappa_2}{3} \int \frac{Ndt}{{\cal U}}
\end{equation}
%\begin{eqnarray}
%\rm a &=& \rm ln \,{\cal U}^{1/3} -\frac{\kappa_2+2\kappa_3}{3} \int %\frac{Ndt}{{\cal U}} \,, \label{solita-a}\\
%\rm b &=& \rm ln \,{\cal U}^{1/3} +\frac{\kappa_2-\kappa_1}{3} \int %\frac{Ndt}{{\cal U}} \,, \label{solita-b}\\
%\rm c &=& \rm ln \,{\cal U}^{1/3} -\frac{\kappa_1+2\kappa_2}{3} \int %\frac{Ndt}{{\cal U}} \,, \label{solita-c}
%\end{eqnarray}
%
therefore the exact solutions will have to meet either eqs.~(\ref{solution-a}-\ref{solution-b-c}) or eqs.~(\ref{solution2-a-b-c}), after finding the functional form of $\rm N(t)$ and $\rm {\cal U}$. In the next section we will employ the Hamiltonian method in order to find the solutions of the whole system.

%%%%%%%%%%%%%%%%%%%%%%%%%%%%%%%%

\section{Hamiltonian approach \label{hap}}

We will implement the Hamiltonian approach to obtain the classical solutions to the EKG eqs.~(\ref{ein0}-\ref{kg-phi}), as well as their counterparts in a quantum scheme. First we need to build the corresponding Lagrangian and Hamiltonian densities for the model in question. We take the metric eq.~(\ref{bianchi-i}) into the Lagrangian density eq.~(\ref{lagra}), having
\begin{equation}
\rm {\cal{L}}= \rm e^{a+b+c}\left(2 \frac{\dot a \dot b}{N}+ 2 \frac{\dot a \dot c}{N}+ 2 \frac{\dot b \dot c}{N}
-\frac{1}{2N} \sum_{i=1}^n (\dot \phi_i)^2+N V(\phi_i,\cdots,\phi_n) \right) \,. \label{lagra-bianchi}
\end{equation}
Then the corresponding momenta are defined in the usual way $\rm \Pi_q=\partial {\cal L}/\partial \dot q$, thus we obtain
\begin{eqnarray}
\rm \Pi_a &=& \rm e^{a+b+c} \frac{2(\dot b + \dot c)}{N} \,,
\qquad\qquad \dot a=\frac{N e^{-(a+b+c)}}{4} \left[-\Pi_a +\Pi_b+ \Pi_c \right] \,, \nonumber\\
\rm \Pi_b &=&\rm e^{a+b+c} \frac{2(\dot a + \dot c)}{N} \,,
\qquad\qquad \dot b=\frac{N e^{-(a+b+c)}}{4} \left[-\Pi_b +\Pi_a+ \Pi_c \right] \,, \nonumber\\
\rm \Pi_c &=&\rm e^{a+b+c} \frac{2(\dot a + \dot b)}{N} \,,
\qquad\qquad \dot c=\frac{N e^{-(a+b+c)}}{4} \left[-\Pi_c +\Pi_a+ \Pi_b \right] \,, \nonumber\\
\rm \Pi_{\phi_i} &=& \rm -\frac{e^{a+b+c}}{N} {\dot \phi_i} ,\qquad\qquad\qquad \,\,
\dot \phi_i = -N e^{-(a+b+c)} \Pi_{\phi_i}, \quad i=1,\cdots,n \,.
\label{momenta}
\end{eqnarray}
By performing the variation of the canonical Lagrangian with respect to $\rm N$, i.e. $\rm\delta{\cal L}_{canonical}/\delta N=0$, where $\rm{\cal L}_{canonical}= \Pi_q \dot{q}-N{\cal H}$, it implies the constraint $\rm{\cal H}=0$. Hence the Hamiltonian density is
\begin{equation}
\rm {\cal H}= \rm \frac{e^{-(a+b+c)}}{8} \left[ - \Pi_a^2 - \Pi_b^2- \Pi_c^2 +2(\Pi_a \Pi_b +\Pi_a \Pi_c + \Pi_b \Pi_c) -4 \sum_{i=1}^n \Pi_{\phi_i}^2
 -8 e^{2(a+b+c)} V(\phi_1,\cdots,\phi_n) \right] \,. \label{hami-bianchi}
\end{equation}
In the gauge $\rm N=8e^{a+b+c}$ and using using the Hamilton equations $\rm \dot{q}=\partial {\cal H}/\partial \Pi_{q}$ and $\rm \dot{\Pi}_q=-\partial {\cal H}/\partial q$, we have the following set of equations
\begin{eqnarray}
\rm \dot a &=& \rm  2\left[ -\Pi_a + \Pi_b + \Pi_c \right],  \qquad\qquad\qquad  \dot \Pi_a = 16 V(\phi_1,\cdots,\phi_n) e^{2(a+b+c)}, \nonumber\\
\rm \dot b &=& \rm  2\left[ -\Pi_b + \Pi_a + \Pi_c \right],  \qquad\qquad\qquad  \dot \Pi_b = 16 V(\phi_1,\cdots,\phi_n) e^{2(a+b+c)}, \nonumber\\
\rm \dot c &=& \rm  2\left[ -\Pi_c + \Pi_a + \Pi_b \right],  \qquad\qquad\qquad  \dot \Pi_c = 16 V(\phi_1,\cdots,\phi_n) e^{2(a+b+c)}, \nonumber\\
\rm \dot \phi_i &=& \rm -8 \Pi_{\phi_i}, \qquad\qquad\qquad\qquad\qquad\qquad \dot \Pi_{\phi_i} = \rm 8 e^{2(a+b+c)} \left(\frac{\partial V(\phi_1,\cdots,\phi_n)}{\partial \phi_i}\right)_{\phi_{j\not=i}}\,, \quad i=1,\cdots,n \,. \label{new-variables}
\end{eqnarray}

To solve the entire system we propose a direct correspondence between the time derivative of the momenta $\dot{\Pi}_{\phi_i} \propto \dot{\Pi}_{\phi_j}$, hence
\begin{equation}
\rm \dfrac{\partial V}{\partial \phi_i} =\alpha_{ij} \dfrac{\partial V}{\partial \phi_j} \,,
\end{equation}
where $\alpha_{ij}$ is the relating constant between the fields i and j. Such connection can be obtained considering two different configurations of the potential: $\rm V(\phi_i,\ldots,\phi_n)=f[\pm(\alpha_1 \phi_i+\cdots+\alpha_n \phi_n)]$, and $\rm V(\phi_i,\ldots,\phi_n)=\sum_{i=1}^n V_i f[\pm(\alpha_i \phi_i)]$, where $\rm f[\pm(\alpha_1 \phi_i+\cdots+\alpha_n \phi_n)]$ is a set of arbitrary functions. The first approach is the simplest one, where such potential can be obtained by the separation variables method. Hence, we  select the straightforward one   
\begin{equation}
\rm V=  V_0 e^{-\left[\lambda_1 \phi_1 + \cdots + \lambda_n \phi_n \right]}\,,
\end{equation}
where $\rm V_{0}$ is a constant and $\rm \{\lambda_{1},\ldots,\lambda_{n}\}$ are n distinguishing parameters. This class of potential has been obtained by other methods, see for instance \cite{gssa,socorro-doleire,socorro-pimentel,ssw,Socorro:2017dle}. Thus the time derivative of i-th momenta becomes 
\begin{equation}
\rm \dot \Pi_{\phi_i}= -8\lambda_ie^{2(a+b+c)} V(\phi_1,\cdots,\phi_n)    
\end{equation}
therefore the momenta are 
\begin{equation}
\rm \Pi_{\phi_i}=-\frac{\lambda_i}{2}\Pi_a-p_{\phi_i}\,,\quad i=1,\cdots,n \,,\qquad
 \qquad  \Pi_b=\Pi_a - p_b \,, \qquad \Pi_c= \Pi_a - p_c, \label{momentos}
\end{equation}
where $\rm p_{\phi_i}\,, p_b$, and $\rm p_c$ are integration constants, which their sign are going to be selected by suitability. In the following sections we will compute and obtain the exact classical and quantum solutions.

\section{Classical solution \label{solutions}}

In this section our goal is to compute the exact classical solutions for this model. Since the variation of the canonical Lagrangian with respect to $\rm N$ implies the constraint $\rm {\cal H}=0$, we thus take it into account to obtain the temporal dependence for $\rm \Pi_a(t)$, hence we construct a master equation
\begin{equation}\label{master-equation}
\rm \frac{d \Pi_a}{m_1 \Pi_a^2 - m_2 \Pi_a - m_3}=dt \,,
\end{equation}
where the parameters $\rm m_i, i=1,2,3 $ are
\begin{eqnarray}
\rm m_1 &=& \rm 2\eta=2\left(3-\sum_{i=1}^n \lambda_i^2\right)=2(3- \lambda^2) \,, \nonumber\\
\rm m_2 &=& \rm 2p_1+8\left[\sum_{i=1}^n \lambda_i p_{\phi_i} \right] \,, \quad p_1=2(p_b+p_c) \,, \label{parameter}\\
\rm m_3 &=& \rm 8 \sum_{i=1}^n p_{\phi_i}^2+ \frac{p_2^2}{2} \,, \quad p_2=2(p_b-p_c) \,. \nonumber
\end{eqnarray}

Subsequently we analyze different cases regarding the parameter $\rm \lambda^2=\sum_{i=1}^n\lambda_i^2$. We will study three distinct scenarios.   

\subsection{Solution for $\lambda^2=3$}

Having $\lambda^2=3$ implies that $\rm m_1=0$, hence the temporal dependence for $\rm \Pi_{a}(t)$ yields by solving the equation
\begin{equation}\label{master-equation3}
\rm \frac{d \Pi_a}{m_2 \Pi_a - m_3}=dt \,,
\end{equation}
%
%for this case the parameters $\rm m_2$ and $\rm m_3$ are
%
%\begin{eqnarray}
%\rm m_2 &=& \rm 2 p_1+ 8\sum_{i=1}^n \lambda_i p_{\phi_i} \,, \qquad p_1=2(p_b+p_c) %\label{parameter3}\\
%\rm m_3 &=& \rm \frac{p_2^2}{2}+8 \sum_{i=1}^n p_{\phi_i}^2 \,, \qquad p_2=2(p_b-p_c) %\,. 
%\end{eqnarray}
%
therefore the solution of eq.~(\ref{master-equation3}) is
\begin{equation}
\rm \Pi_a(t)=\frac{m_3}{m_2} +c_1 e^{m_2 t},\label{master}
\end{equation}
where $\rm c_1$ is an integration constant. From the set of equations~(\ref{new-variables}), we obtain the corresponding solutions for the set of variables $\rm(a,b,c,\phi_i)$ and
$\rm (\Pi_a, \Pi_b, \Pi_c, \Pi_{\phi_i})$, which are
\begin{eqnarray}
\rm a= a_0 +\left(\frac{2m_3}{m_2} + p_1 \right)t + \frac{2c_1}{m_2} e^{m_2t}, \nonumber\\
\rm b = \rm b_0 +\left(\frac{2m_3}{m_2} + p_2 \right)t + \frac{2c_1}{m_2} e^{m_2t} \,, &&\quad\rm \Pi_b(t)= \left(\frac{m_3}{m_2} + p_b \right) +c_1 e^{m_2 t} \,, \nonumber\\
\rm c = \rm c_0 +\left(\frac{2m_3}{m_2} - p_2 \right)t + \frac{2c_1}{m_2} e^{m_2t} \,, &&\quad\rm  \Pi_c(t)= \left(\frac{m_3}{m_2} + p_c \right) +c_1 e^{m_2 t}\,, \nonumber\\
\rm \phi_i = {\phi_i}_0  +\left(\frac{2\lambda_i m_3}{m_2} 8 p_{\phi_i} \right)t+ \frac{4 \lambda_i c_1}{m_2} e^{m_2t} \,, && \quad\rm  \Pi_{\phi_i} = \rm \left( -\frac{\lambda_i m_3}{2 m_2}+ p_{\phi_i} \right) -\frac{\lambda_i c_1}{2}  e^{m_2t} \,,\quad i=1,\cdots,n \,, \label{solutions-lambda3}
\end{eqnarray}
%
%\begin{eqnarray}
%\rm a &=& \rm a_0 +\left(\frac{2m_3}{m_2} + p_1 \right)t + \frac{2c_1}{m_2} e^{m_2t}, \nonumber\\
%\rm b &=& \rm b_0 +\left(\frac{2m_3}{m_2} + p_2 \right)t + \frac{2c_1}{m_2} e^{m_2t}, \nonumber\\
%\rm c &=& \rm c_0 +\left(\frac{2m_3}{m_2} - p_2 \right)t + \frac{2c_1}{m_2} e^{m_2t}, \nonumber\\
%\rm \phi_i &=& \rm {\phi_i}_0  +\left(\frac{2\lambda_i m_3}{m_2} 8 p_{\phi_i} \right)t+ \frac{4 \lambda_i c_1}{m_2} e^{m_2t}, \qquad i=1,\cdots,n \\ \label{solutions-lambda3}
%\rm \Pi_b(t)&=& \rm \left(\frac{m_3}{m_2} + p_b \right) +c_1 e^{m_2 t},\nonumber\\
%\rm \Pi_c(t)&=& \rm \left(\frac{m_3}{m_2} + p_c \right) +c_1 e^{m_2 t},\nonumber\\
%\rm \Pi_{\phi_i} &=& \rm \left( -\frac{\lambda_i m_3}{2 m_2}+ p_{\phi_i} \right) -\frac{\lambda_i c_1}{2}  e^{m_2t},\quad i=1,\cdots,n, \nonumber
%\end{eqnarray}
%
where $\rm a_0, b_0, c_0, {\phi_i}_0$ are integration constants. Notice that we have selected as positive the constants $\rm +p_{b},+p_{c}$. In order for the above results to fulfill the EKG eqs.~(\ref{ein0}-\ref{kg-phi}), all constants must satisfy $\rm 16 V_0=m_2 c_1 Exp\left[-2a_0-2b_0-2c_0+ \sum_{i=1}^n\lambda_i {\phi_{i\,0}}\right]$. Moreover, by evaluating the solutions eqs.~(\ref{solution2-a-b-c}), the parameters $\rm \kappa_i$ follow the relations: 
\begin{equation}
\rm \kappa_1= \frac{p_c}{2} \,, \qquad \kappa_2= \frac{p_b-p_c}{2} \,, \qquad \kappa_3=-\frac{p_b}{2} \,.   
\end{equation}
Finally, the scale factors for this scenario become
\begin{eqnarray}
\rm  A &=& \rm A_0 Exp\left[\left(\frac{2m_3}{m_2} + p_1 \right)t  \right]  Exp\left[ \frac{2c_1}{m_2} e^{m_2t}\right] \,,\quad A_{0}=e^{a_{0}} \,,\nonumber\\
\rm  B &=& \rm B_0 Exp\left[\left(\frac{2m_3}{m_2} + p_2 \right)t  \right]  Exp\left[ \frac{2c_1}{m_2} e^{m_2t}\right] \,,\quad B_{0}=e^{b_{0}} \,,\\ \label{scale-factor-lambda3}
\rm  C &=& \rm C_0 Exp\left[\left(\frac{2m_3}{m_2} - p_2 \right)t  \right]  Exp\left[ \frac{2c_1}{m_2} e^{m_2t}\right] \,,\quad C_{0}=e^{c_{0}} \,.\nonumber
\end{eqnarray}

\subsection{Solution for $\lambda^2 < 3$}

For this case the solution of eq.~(\ref{master-equation}) becomes
\begin{equation}
\rm -\Delta t=\frac{1}{\omega} Ln\left( \frac{-4\eta \Pi_a + b-\omega}{p_1(-4\eta \Pi_a + b+\omega)}
\right) \,,    
\end{equation}
where $\rm \omega^2=m_2^2+8\eta m_3$, therefore $\Pi_a$ is
\begin{equation}
\rm \Pi_a = \frac{1}{4\eta}\left[m_2-\omega\coth\left(\frac{\omega}{2}t\right)\right] \,.
\end{equation}
Once again, from the set of equations~(\ref{new-variables}), the corresponding solutions for our set of variables are
\begin{eqnarray}
\rm a= a_0 +\left( \frac{m_2}{2\eta}-p_1\right) t- \frac{1}{\eta} \ln{\left[sinh{\left(\frac{\omega}{2}t\right)}\right]} \,,\nonumber\\
\rm b=b_0 + \left( \frac{m_2}{2\eta}+p_2\right)t- \frac{1}{\eta} \ln{\left[sinh{\left(\frac{\omega}{2}t\right)}\right]} \,, &&\quad\rm \Pi_b= \frac{1}{4\eta} \left[m_2-\omega\coth \left(\frac{\omega}{2}t\right)\right]-p_b \,,\nonumber\\
\rm c = c_0 + \left( \frac{m_2}{2\eta}-p_2\right)t- \frac{1}{\eta} \ln{\left[sinh{\left(\frac{\omega}{2}t\right)}\right]} \,, &&\quad \rm \Pi_c =  \frac{1}{4\eta} \left[m_2-\omega\coth \left(\frac{\omega}{2}t\right)\right]-p_c \,,\nonumber\\
\hspace{-0.5cm}\rm \phi_i={\phi_i}_0 + \left(\lambda_i \frac{m_2}{\eta}+8p_{\phi_i}\right) t -
\frac{2 \lambda_i}{\eta}\, \ln{\left[\sinh{\left(\frac{\omega}{2}t\right)}\right]} \,, &&\quad \rm \Pi_{\phi_i} = -\frac{\lambda_i}{8\eta}\left[m_2-\omega coth \left(\frac{\omega}{2}t\right)\right] -p_{\phi_i}\,,\,\, i=1,\cdots,n \,, \label{solutions-lambda-less3}
\end{eqnarray}
%
%\begin{eqnarray}
%\rm a &=& \rm a_0 +\left( \frac{m_2}{2\eta}-p_1\right) t- \frac{1}{\eta} \ln{\left[sinh{\left(\frac{\omega}{2}t\right)}\right]} \,,\nonumber\\
%\rm b &=& \rm b_0 + \left( \frac{m_2}{2\eta}+p_2\right)t- \frac{1}{\eta} \ln{\left[sinh{\left(\frac{\omega}{2}t\right)}\right]} \,,\nonumber\\
%\rm c &=& \rm c_0 + \left( \frac{m_2}{2\eta}-p_2\right)t- \frac{1}{\eta} \ln{\left[sinh{\left(\frac{\omega}{2}t\right)}\right]} \,,\nonumber\\
%\rm \phi_i &=& \rm {\phi_i}_0 + \left(\lambda_i \frac{m_2}{\eta}+8p_{\phi_i}\right) t -
%\frac{2 \lambda_i}{\eta}\, \ln{\left[\sinh{\left(\frac{\omega}{2}t\right)}\right]}, \quad i=1,\cdots,n \,,\\
%\rm \Pi_b &=& \rm \frac{1}{4\eta} \left[m_2-\omega\coth \left(\frac{\omega}{2}t\right)\right]-p_b \,, \nonumber\\
%\rm \Pi_c &=& \rm \frac{1}{4\eta} \left[m_2-\omega\coth \left(\frac{\omega}{2}t\right)\right]-p_c \,, \nonumber\\
%\rm \Pi_{\phi_i} &=& \rm   -\frac{\lambda_i}{8\eta}\left[m_2-\omega coth \left(\frac{\omega}{2}t\right)\right] -p_{\phi_i} \,,\nonumber
%\end{eqnarray}
where $\rm a_0, b_0, c_0, {\phi_i}_0$ are integration constants. Notice that this time we have selected as negative the constants $\rm -p_{b},-p_{c}$. In order for the above results to fulfill the EKG eqs.~(\ref{ein0}-\ref{kg-phi}), all constants must satisfy $\rm 1152 \eta V_0=\omega^2 Exp\left[-2a_0-2b_0-2c_0+\sum_{i=1}^n\lambda_i {\phi_i}_0\right]$. Additionally, by evaluating the solutions eqs.~(\ref{solution2-a-b-c}), the parameters $\rm \kappa_i$ follow the relations:
\begin{equation} 
\rm \kappa_1= -\frac{p_c}{2}, \qquad \kappa_2= \frac{2p_c-p_b}{4}, \qquad \kappa_3=\frac{p_b}{4} \,.    
\end{equation}
Finally, the scale factors for this case become
\begin{eqnarray}
\rm A &=& \rm A_0 Exp\left[ \left(\frac{m_2}{2\eta}-p_1\right) t\right] \,
csch^{1/\eta} \left(\frac{\omega}{2}t\right) \,, \nonumber\quad A_{0}=e^{a_{0}}\,,\\
\rm B &=& \rm B_0 Exp\left[ \left( \frac{m_2}{2\eta}+p_2\right) t\right] \,
csch^{1/\eta} \left(\frac{\omega}{2}t\right) \,, \quad B_{0}=e^{b_{0}}\,,\\
\label{scale-factor-lambdaless3}
\rm C &=& \rm C_0 Exp\left[ \left( \frac{m_2}{2\eta}-p_2\right) t\right] \,
csch^{1/\eta} \left(\frac{\omega}{2}t\right) \,, \nonumber\quad C_{0}=e^{c_{0}} \,. 
\end{eqnarray}

\subsection {Solution for $\lambda^2>3$}

For this case, notice that the parameter $\rm m_1=2(3-\lambda^2)<0$, so we introduce $\beta=(\lambda^2-3)>0$ 
\begin{equation}\label{master-integral-lambda-bigger3}
\rm \frac{d \Pi_a}{-2\beta\Pi_a^2-m_2 \Pi_a - m_3}=dt \,,
\end{equation}
therefore the solution to the momenta $\rm \Pi_a(t)$ becomes
\begin{equation}
\rm \Pi_a=\frac{m_2}{4\beta}+\frac{\omega}{4\beta} tanh\left( \frac{\omega}{2}t \right) \,.
\end{equation}
provided that $\rm m_2^2 -8\beta m_3>0$ then $\rm \omega^2 =m_2^2 -8\beta m_3$. From the set of equations~(\ref{new-variables}), the corresponding solutions for our set of variables are 
\begin{eqnarray}
\rm a=a_0 + \left(\frac{m_2}{2\beta}+p_1\right) t + \frac{1}{\beta} Ln\left[cosh\left(\frac{\omega}{2}t\right)\right]  \,,\nonumber\\
\rm b=b_0 + \left( \frac{m_2}{2\beta}-p_2\right)t+ \frac{1}{\beta} \ln{\left[cosh{\left(\frac{\omega}{2}t\right)}\right]} \,,&&\quad\rm \Pi_b=\frac{m_2}{4\beta}+\frac{\omega}{4\beta} tanh\left( \frac{\omega}{2}t \right)+p_b \,,\nonumber\\
\rm c = \rm c_0 + \left( \frac{m_2}{2\beta}+p_2\right)t+ \frac{1}{\beta} \ln{\left[cosh{\left(\frac{\omega}{2}t\right)}\right]} \,,&&\quad \rm \Pi_c=\frac{m_2}{4\beta}+\frac{\omega}{4\beta} tanh\left( \frac{\omega}{2}t \right)+p_c \,,\nonumber\\
\hspace{-0.4cm}\rm \phi_i = \rm {\phi_i}_0 + \left(\lambda_i \frac{m_2}{\beta}-8p_{\phi_i}\right) t +
\frac{2 \lambda_i}{\beta}\, \ln{\left[\cosh{\left(\frac{\omega}{2}t\right)}\right]} \,,&&\quad \rm \Pi_{\phi_i}=-\frac{\lambda_{i}}{2}\left(\frac{m_2}{4\beta}+\frac{\omega}{4\beta} tanh\left( \frac{\omega}{2}t \right)\right)+p_{\phi_i} \,,\,\, i=1,\ldots,n \,,
\end{eqnarray}
%
%\begin{eqnarray}
%&&\rm a=a_0 + \left(\frac{m_2}{2\beta}+p_1\right) t + \frac{1}{\beta} Ln\left[cosh\left(\frac{\omega}{2}t\right)\right]  \,,\nonumber\\
%&&\rm b=b_0 + \left( \frac{m_2}{2\beta}-p_2\right)t+ \frac{1}{\beta} \ln{\left[cosh{\left(\frac{\omega}{2}t\right)}\right]} \,,\nonumber\\
%&&\rm c = \rm c_0 + \left( \frac{m_2}{2\beta}+p_2\right)t+ \frac{1}{\beta} \ln{\left[cosh{\left(\frac{\omega}{2}t\right)}\right]} \,,\nonumber\\
%&&\rm \phi_j = \rm {\phi_j}_0 + \left(\lambda_j \frac{m_2}{\beta}-8p_{\phi_j}\right) t +
%\frac{2 \lambda_j}{\beta}\, \ln{\left[\cosh{\left(\frac{\omega}{2}t\right)}\right]} \,, \quad j=1,\cdots,n \,,\\
%&&\rm \Pi_b=\frac{m_2}{4\beta}+\frac{\omega}{4\beta} tanh\left( \frac{\omega}{2}t \right)+p_b \,,\nonumber\\
%&&\rm \Pi_c=\frac{m_2}{4\beta}+\frac{\omega}{4\beta} tanh\left( \frac{\omega}{2}t \right)+p_c \,,\nonumber\\
%&&\rm \Pi_{\phi_j}=-\frac{\lambda_{j}}{2}\left(\frac{m_2}{4\beta}+\frac{\omega}{4\beta} tanh\left( \frac{\omega}{2}t \right)\right)+p_{\phi_j} \,,\nonumber
%\end{eqnarray}
%
where $\rm a_0, b_0, c_0, \phi_{j0}$ are all integration constants. This time we have selected as positive the constants $\rm p_{b},p_{c}$. In order for the above results to fulfill the EKG eqs.~(\ref{ein0}-\ref{kg-phi}), all constants must satisfy $\rm 128 \beta V_0=\omega^{2}Exp\left[-2a_0-2b_0-2c_0+\sum_{i=1}^n\lambda_i {\phi_i}_0\right]$. Additionally, by evaluating the solutions eqs.~(\ref{solution2-a-b-c}), the parameters $\rm \kappa_i$ follow the relations:
\begin{equation}
\rm \kappa_1= \frac{p_b}{2}, \qquad \kappa_2= \frac{p_c-p_b}{2}, \qquad \kappa_3=-\frac{p_c}{2} \,,    
\end{equation}
Finally, the scale factors for this framework become
\begin{eqnarray}
\rm A &=& \rm A_0 Exp\left[ \left(\frac{m_2}{2\beta}+p_1\right) t\right] \,
cosh^{1/\beta} \left(\frac{\omega}{2}t\right) \,, \quad A_{0}=e^{a_{0}} \,,\nonumber\\
\rm B &=& \rm B_0 Exp\left[ \left( \frac{m_2}{2\beta}-p_2\right) t\right] \,
cosh^{1/\beta} \left(\frac{\omega}{2}t\right) \,, \quad B_{0}=e^{b_{0}} \,,\\
\rm C &=& \rm C_0 Exp\left[ \left( \frac{m_2}{2\beta}+p_2\right) t\right] \,
cosh^{1/\beta} \left(\frac{\omega}{2}t\right) \,, \quad C_{0}=e^{c_{0}} \,. \label{scale-factorslambdalarger3}
\end{eqnarray}

\subsection{Anisotropic density}

We can measure the anisotropic density by implementing the Misner's parameterization $(\Omega, \beta_+, \beta_-)\leftrightarrow (a,b,c)$, where $(\beta_+,\beta_-)$ are the anisotropic parameters:
\begin{eqnarray}
\rm a &=& \rm  \Omega+\beta_+\sqrt{3} \beta_- \,, \qquad b= \rm \Omega+\beta_+-\sqrt{3} \beta_- \,,\qquad c= \rm \Omega-2\beta_+ \,,\nonumber\\
\rm \Omega&=& \rm \frac{1}{3}(a+b+c) \,,\qquad  \beta_+=\frac{1}{6}(a+b-2c) \,, \qquad\quad \beta_-=\frac{\sqrt{3}}{6}(a-b) \,.
\end{eqnarray}
The anisotropic and gravitational densities are defined by $\rm\rho_{anisotropic}=(\dot \beta_+)^2 + (\dot \beta_-)^2$ and $\rho_\Omega=(\dot \Omega)^2$, respectively. When the anisotropic-to-gravitational density rate goes to zero the spacetime becomes isotropic \cite{jsocorro}. Remarkably, in all cases the anisotropic density is 
\begin{equation}
\rm \rho_{anisotropic}= \frac{p_1^2+3 p_2^2}{9}=constant \,,    
\end{equation}
since $\rm\dot{\Omega}$ increases with time, isotropization is indeed reached eventually.

\subsection{Number of e-folds}

The Bianchi type I model becomes the FRLW spacetime when the three scale factors are equal; for this reason it is convenient to introduce the average scale factor, defined by
\begin{equation}\label{average-scale-factor}
\rm S(t)=(ABC)^{1/3}=e^{(a+b+c)/3} \,,    
\end{equation}
which characterizes the volume expansion of the universe, and for a particular case $\rm A=B=C$ one recovers the scale factor described in the FRLW metric. Then, we introduce the associated Hubble parameter
\begin{equation}\label{Hubble}
\rm H(t)=\frac{\dot{S}}{S}=\frac{\dot{a}+\dot{b}+\dot{c}}{3} \,,
\end{equation}
and since the physical time is $\rm dt_{phys}=Ndt$, the physical Hubble parameter becomes:
\begin{equation}\label{Hubble-physical}
\rm H_{phys}=\frac{H(t)}{N(t)}=\frac{\dot{a}(t)+\dot{b}(t)+\dot{c}(t)}{3N(t)}=\frac{1}{3}\left(\frac{da(t_{phys})}{dt_{phys}}+\frac{db(t_{phys})}{dt_{phys}}+\frac{dc(t_{phys})}{dt_{phys}} \right) \,.
\end{equation}
Inflation is characterized by the number of e-folds it expands to during such period, that corresponds to $\rm S''_{phys}>0$, where the primes denote the derivatives with respect to the cosmic time $\rm t_{phys}$. The e-folding function 
\begin{equation}
\rm N_{e}=\int_{t_{phys}*}^{t_{phys\,\,end}} dt_{phys}H(t_{phys})=\int_{t*}^{t_{end}} dt\frac{N}{N}H=\int_{t*}^{t_{end}} \frac{dt}{3}(\dot{a}+\dot{b}+\dot{c}) \,, 
\end{equation}
where $\rm t_{phys}*$ represents the time when the relevant cosmic microwave background (CMB) modes become superhorizon at 50-60 e-folds before inflation ends at $\rm t_{phys\,\,end}$; when in the proper time $\rm t$ the equivalents are $\rm t*$ and $\rm t_{end}$, respectively. At the end of inflation the second derivative of the average scale factor must satisfy $\rm S''_{phys}=0$, or the condition $\rm -dH_{phys}/dt_{phys}=H_{phys}^{2}$ must be met, which in our gauge becomes      
\begin{equation}\label{condition-inflation-end}
\rm \ddot{a}+\ddot{b}+\ddot{c}=\frac{2}{3}(\dot{a}+\dot{b}+\dot{c})^{2} \,,     
\end{equation}
therefore one can compute the period when inflation finalizes by solving eq.~(\ref{condition-inflation-end}) for each scenario. In Table \ref{t:solutions} appears the computation of the e-folding function $\rm N_{e}$ and $\rm t_{end}$ for each case given by the $\rm\lambda$ parameter. 
\begin{center}
\begin{table}[ht]
\begin{scriptsize}
\renewcommand{\arraystretch}{3.5}
\begin{tabular}{|r|c|c|c|}
\hline
 & $\rm t_{end}$ & $\rm M_{\pm}^{(n)}$  &$\rm N_{e}$   \\ \hline
$\lambda^{2}=3$ & $\rm\frac{1}{m_{2}}\ln\left(\frac{M_{\pm}^{(1)}}{c_{1}}\right)$ & $\rm \frac{1}{24}\left[3m_{2}-\frac{24m_{3}}{m_{2}}-4p_{1}\pm\sqrt{3}\sqrt{3m_{2}^{2}-48m_{3}-8m_{2}p_{1}}\right] $ & $\rm 2\left(\frac{m_{3}}{m_{2}}+\frac{p_{1}}{6}\right)(t_{end}-t_{*})+\frac{2c_{1}}{m_{2}}\left( e^{m_{2}t_{end}}-e^{m_{2}t_{*}} \right)$  \\ \hline
$\lambda^{2}<3$ & $\rm \frac{2}{\omega}arccoth\left(M^{(2)}_{\pm}\right)$ &$\rm\frac{1}{3\omega(2-\eta)}\left[6m_{2}-4\eta p_{1}\pm\sqrt{\eta(9(-2+\eta) \omega^{2}+2(3m_{2}-2\eta p_{1})^{2})}\right] $& $\rm\left(\frac{m_{2}}{2\eta}-\frac{p_{1}}{3}\right)(t_{end}-t_{*})-\frac{1}{\eta}\ln\left[\frac{\sinh\left(\frac{\omega}{2}t_{end}\right)}{\sinh\left(\frac{\omega}{2}t_{*}\right)}\right] $ \\ \hline
$\lambda^{2}>3$ & $\rm\frac{2}{\omega}arctanh\left(M^{(3)}_{\pm}\right)$ &$\rm-\frac{1}{3\omega(2+\beta)}\left[6m_{2}+4\beta p_{1}\pm\sqrt{\beta(9(2+\beta) \omega^{2}-2(3m_{2}+2\beta p_{1})^{2})}\right]$& $\rm\left(\frac{m_{2}}{2\beta}+\frac{p_{1}}{3}\right)(t_{end}-t_{*})+\frac{1}{\beta}\ln\left[\frac{\cosh\left(\frac{\omega}{2}t_{end}\right)}{\cosh\left(\frac{\omega}{2}t_{*}\right)}\right]$\\    
\hline
\end{tabular}
\renewcommand{\arraystretch}{1}
\end{scriptsize}
\caption{Computation of the number of e-folds $\rm N_{e}$ and $\rm t_{end}$ for each case provided by the $\rm\lambda$ parameter. Note that $\rm M_{\pm}^{(1)}>1$, $\rm |M_{\pm}^{(2)}|>1$, and $\rm |M_{\pm}^{(3)}|<1$ in order to have that $\rm t_{end}>0$.  \label{t:solutions} }
\end{table}
\end{center}
We want to illustrate with a simple case, for $\rm\lambda^2=3$, that a particular parameter space can indeed drive at least 60 e-folds of inflation. Hence, we consider $\rm p_{1}=(3m_{2}^{2}-48m_{3})/(8m_{2})$, it yields:
\begin{equation}
\rm N_{e}=\frac{1}{8}\ln\left(\frac{m_{2}}{16c_{1}}\right)+\frac{1}{8}-\frac{m_{2}}{8}t_{*}-\frac{2c_{1}}{m_{2}}e^{m_{2}t_{*}} \,,\qquad c_{1}<\frac{m_{2}}{16} \,.   
\end{equation}
From the above equation one can immediately notice that in order to obtain 60 e-folds the logarithm function must dominate, hence we may take $\rm m_{2}/c_{1}\sim 48\times 10^{208}$ and $\rm m_{2}t_{*}\sim\mathcal{O}(1)$, which in turn makes the factor in front of the exponential to become $\rm c_{1}/m_{2}\sim 10^{-208}\sim 0$. The aforementioned implies that  $\rm t_{end}$ will be the same order of magnitude as $\rm t_{*}$ since from the Table \ref{t:solutions} we can see that $\rm t_{end} \sim 1/m_2$, which indeed fulfills $\rm t_{end}>t_{*}$. The remaining cases are not fully studied, since it requires a thoroughly computational numerical analysis, however, such exhaustive inspection was outside the scope of this paper.

%%%%%%%%%%%%%%%%%%%%%%%%%%%%%%%%%%%%%%%%%%%%%%%%%%%%%%%%%
%%%%%%%%%%%%%%%%%%%%%%%%%%%%%%%%%%%%%%%%%%%%%%%%%%%%%%%%%
%%%%%%%%%%%%%%%%%%%%%%%%%%%%%%%%%%%%%%%%%%%%%%%%%%%%%%%%%

\section{Quantum solution\label{quantum}}

In order to simplify the mathematical apparatus in the multi-scalar field quantum scheme, we study it considering only two scalar constituents ($\phi, \sigma$). We begin with the classical Hamiltonian density
\begin{equation}
\rm {\cal H}= \rm \frac{e^{-(a+b+c)}}{8} \left[ - \Pi_a^2 - \Pi_b^2- \Pi_c^2 +2(\Pi_a \Pi_b +\Pi_a \Pi_c + \Pi_b \Pi_c) -4 \Pi_\phi^2
-4 \Pi_\sigma^2 -8V_0 e^{2(a+b+c)-\lambda_1 \phi -\lambda_2 \sigma}  \right] \,. \label{hami-bianchi-quantico}
\end{equation}
We introduce a new set of variables: 
\begin{eqnarray}
\rm\xi &=& \rm  a+b+\sqrt{3} c \,, \qquad \eta= \rm a+b-\sqrt{3} c \,,\qquad \chi= \rm a-2b \,,\nonumber\\
\rm\rho &=& \rm \alpha_1 a+ \alpha_2 b+\alpha_3 c +\alpha_4 \phi+ \alpha_5 \sigma \,, \qquad \rm\nu=\rm a + b+ c-\frac{\lambda_1}{2}\phi -\frac{\lambda_2}{2}\sigma \,, \label{new-coordenadas}
\end{eqnarray}
where $\rm\{\lambda_{1},\lambda_{2}\}$ are the parameters of the exponential potential, $\rm\{\alpha_{1},\alpha_{2},\alpha_{3},\alpha_{4},\alpha_{5}\}$ are free parameters. Thus, the inverse transformations are:
\begin{eqnarray}
\rm a &=& \rm \frac{1}{3}(\xi+\eta+\chi),\qquad  b=\frac{1}{6}(\xi+\eta-2\chi) \,, \qquad\quad c=\frac{\sqrt{3}}{6}(\xi-\eta)\nonumber\\
\rm\phi&=&\rm \frac{1}{6\sqrt{3}(\alpha_4 \lambda_2-\alpha_5 \lambda_1)}\left\{ -\left[\sqrt{3}\lambda_2(2\alpha_1+\alpha_2)+6\alpha_5(\sqrt{3}+1)+3\alpha_3 \lambda_2\right]\xi  \right. \nonumber\\ && \left. + \left[-\sqrt{3}\lambda_2(2\alpha_1+\alpha_2)-6\alpha_5(\sqrt{3}-1)+3\alpha_3 \lambda_2\right]\eta+  2\sqrt{3} \lambda_2(\alpha_2-\alpha_1) \chi +6\sqrt{3}\lambda_2 \rho+12\sqrt{3}\alpha_5 \nu \right\} \,,\nonumber\\
\rm\sigma&=&\rm \frac{1}{6\sqrt{3}(\alpha_4 \lambda_2-\alpha_5 \lambda_1)}\left\{ \left[\sqrt{3}\lambda_1(2\alpha_1+\alpha_2)+6\alpha_4(\sqrt{3}+1)+3\alpha_3 \lambda_1\right]\xi \right. \nonumber\\ && \left. + \left[\sqrt{3}\lambda_1(2\alpha_1+\alpha_2)+6\alpha_4(\sqrt{3}-1)-3\alpha_3 \lambda_1\right]\eta+  2\sqrt{3} \lambda_1(\alpha_1-\alpha_2) \chi -6\sqrt{3}\lambda_1 \rho-12\sqrt{3}\alpha_4 \nu \right\} \,,
 \end{eqnarray}
and the momenta:
\begin{eqnarray}
\rm \Pi_a&=& \rm P_\xi +P_\eta + P_\chi +\alpha_1 P_\rho+ P_\nu \,, \nonumber\\
\rm \Pi_b&=& \rm P_\xi + P_\eta -2 P_\chi+ \alpha_2 P_\rho+ P_\nu \,,\nonumber\\
\rm \Pi_c&=& \rm \sqrt{3}P_\xi -\sqrt{3} P_\eta +\alpha_3 P_\rho+ P_\nu \,, \nonumber\\
\rm \Pi_\phi&=& \rm \alpha_4P_\rho -\frac{\lambda_1}{2} P_\nu \,, \nonumber\\
\rm \Pi_\sigma&=& \rm \alpha_5P_\rho  -\frac{\lambda_2}{2} P_\nu \,, \label{new-momentos}
\end{eqnarray}
where $\rm\{ P_\xi, P_\eta, P_\chi, P_\rho, P_\nu\}$ are the momenta associated with the new variables. In the gauge $\rm N=8e^{a+b+c}$, the Hamiltonian density written in terms of the new set of variables becomes 
\begin{eqnarray}
&& (3-\lambda_1^2 +\lambda_2^2)P_\nu^2 + \left[-(\alpha_1-\alpha_2)^2 +2\alpha_3 ( \alpha_1 + \alpha_2) -\alpha_3^2-4(\alpha_{4}^2+\alpha_5^2)\right]P_\rho^2 -9P_\chi^2
-(4\sqrt{3}+3)P_\eta^2 +(4\sqrt{3}-3)P_\xi^2\nonumber \\
&&\rm + 6 P_\xi P_\eta - 2\sqrt{3}P_\xi  P_\chi  +2\left[ \sqrt{3}(\alpha_1+\alpha_2)+\alpha_3(2-\sqrt{3}) \right]P_\xi P_\rho  +2(\sqrt{3}+2)P_\xi P_\nu +  2\sqrt{3}P_\eta P_\chi \nonumber\\
&& \rm +2\left[-\sqrt{3}(\alpha_1+\alpha_2)+\alpha_3(2+\sqrt{3}) \right]P_\eta P_\rho
+2(-\sqrt{3}+2)P_\eta P_\nu +2\left[3(\alpha_2-\alpha_1)-\alpha_3\right]P_\chi P_\rho  -2P_\chi P_\nu \nonumber\\
&&\rm +  2\left[\alpha_1+\alpha_2+\alpha_3+
2\alpha_4 \lambda_1 +2 \alpha_5 \lambda_2  \right] P_\rho P_\nu   - 8V_0 e^{2\nu}=0 \,. 
\label{hami-bianchi-new}
\end{eqnarray}
The WDW equation for this model is obtained by replacing $\rm P_{q}=-i\hbar \partial_{q}$ in eq.~(\ref{hami-bianchi-new}), and by applying to the wave function $\rm\Psi(\xi,\eta,\chi,\rho,\nu)$, yielding 
\begin{eqnarray}
&& -a_0 \frac{\partial^2 \Psi}{\partial \nu^2} - a_1  \frac{\partial^2 \Psi}{\partial \rho^2} +9 \frac{\partial^2 \Psi}{\partial \chi^2}
+  a_2 \frac{\partial^2 \Psi}{\partial \eta^2}
 - a_3   \frac{\partial^2 \Psi}{\partial \xi^2}  - 6  \frac{\partial^2 \Psi}{\partial \xi \partial \eta}  + 2\sqrt{3}  \frac{\partial^2 \Psi}{\partial \xi \partial \chi}    - a_4 \frac{\partial^2 \Psi}{\partial \xi \partial \rho}  -a_5 \frac{\partial^2 \Psi}{\partial \xi \partial \nu}    -  2\sqrt{3} \frac{\partial^2 \Psi}{\partial \eta \partial \chi} -a_6  \frac{\partial^2 \Psi}{\partial \eta \partial \rho}
 \nonumber\\
&&\rm - a_7   \frac{\partial^2 \Psi}{\partial \eta \partial \nu} - a_8
\frac{\partial^2 \Psi}{\partial \chi \partial \rho} + 2 \frac{\partial^2 \Psi}{\partial \chi \partial \nu}  -  a_9  \frac{\partial^2 \Psi}{\partial \rho \partial \nu}   - \frac{8V_0}{\hbar^2} e^{2\nu} \Psi=0 \,, \label{wdw}
 \end{eqnarray}
where the constants $\rm a_i$ are
\begin{eqnarray}
\rm a_0=3-\lambda_1^2-\lambda_2^2 \,, \qquad a_1 &=& \left[-(\alpha_1-\alpha_2)^2 +2\alpha_3 ( \alpha_1 + \alpha_2) -\alpha_3^2-4(\alpha_{4}^2+\alpha_5^2)\right] \,, \nonumber\\
\rm a_2=(4\sqrt{3}+3) \,, \qquad a_3 &=&  (4\sqrt{3}-3) \,, \nonumber\\
\rm  a_4=2 \left[ \sqrt{3}(\alpha_1+\alpha_2)+\alpha_3(2-\sqrt{3}) \right] \,, \qquad a_5 &=& 2(\sqrt{3}+2) \,,\\
\rm  a_6= 2  \left[-\sqrt{3}(\alpha_1+\alpha_2)+\alpha_3(2+\sqrt{3}) \right]\,,\qquad a_7&=& 2(2-\sqrt{3}) \,, \nonumber\\
\rm a_8=2\left[3(\alpha_2-\alpha_1)-\alpha_3\right] \,, \qquad  a_9 &=& 2 \left[\alpha_1+\alpha_2+\alpha_3+ 2\alpha_4 \lambda_1 +2 \alpha_5 \lambda_2  \right] \,.\nonumber 
\end{eqnarray}
Due to the algebraic structure of eq.~(\ref{wdw}), we introduce the following ansatz for the wave function $\rm \Psi=e^{\frac{1}{\hbar}\left(\delta_1 \xi + \delta_2 \eta + \delta_3 \chi + \delta_4 \rho\right)} \, G(\nu)$, where $\delta_i$ are constants and $\rm G(\nu)$ is the function to be determined. Thus, we obtain an equation for $\rm G(\nu)$ by replacing the ansatz into eq.~(\ref{wdw}), yielding
\begin{equation}
\rm a_0\frac{d^2G}{d\nu^2} + \frac{b_1}{\hbar }\frac{dG}{d\nu} + \left(\frac{8V_0}{ \hbar^2} e^{2\nu}- \frac{b_2}{\hbar^2 }\right)G=0 \,,
\label{mast}
\end{equation}
where
\begin{eqnarray}
\rm b_1 &=&\rm 2 \left[\delta_1(\sqrt{3}+2) + \delta_2(2-\sqrt{3})-\delta_{3}+(\alpha_1+\alpha_2+\alpha_3)\delta_4+2\delta_4(\alpha_4 \lambda_1+\alpha_5 \lambda_2)\right] \,, \nonumber\\
\rm b_2 &=&\rm 4\sqrt{3}(\delta_2^2 - \delta_1^2)+2\sqrt{3}\delta_{3}(\delta_1-\delta_2)+ \delta_4^2(\alpha_1-\alpha_2)^2 +3(\delta_1-\delta_2)^2 + 9\*\delta_{3}^2+2\delta_4 \alpha_3(\delta_3-2\delta_1)\nonumber\\
&& - 2\sqrt{3}\delta_1 \delta_4(\alpha_3-\alpha_1-\alpha_2)+\delta_4^2\left[4(\alpha_4^2+\alpha_5^2)+\alpha_3^2-2\alpha_2\alpha_3-2\alpha_1 \alpha_3\right] \,. 
\end{eqnarray}
Then the solution to the function G becomes \cite{polyanin}:
\begin{equation}\label{G-solution-generic}
\rm G(\nu)= e^{- \frac{a_1}{2\hbar a_0}\nu} Z_\mu\left(\frac{2}{\hbar}\sqrt{\frac{2V_0}{a_0}} e^{\nu}\right)\,,\qquad \mu=\frac{\sqrt{b_{1}^{2}+4a_{0}b_{2}}}{2\hbar a_{0}}
\end{equation}
where $\rm Z_\mu$ is a general Bessel function and $\rm\mu$ is the corresponding order. Note that above eq.~(\ref{G-solution-generic}) only allows solutions for $\rm a_{0}\neq 0$. Thus, the wave function becomes  
\begin{eqnarray}
\rm \Psi&=& \rm \psi_0 e^{\frac{1}{\hbar}\left(\delta_1 \xi + \delta_2 \eta + \delta_3 \chi + \delta_4 \rho +\frac{b_1}{2|a_0|}\nu\right)} K_\mu\left(\frac{2}{\hbar}\sqrt{\frac{2V_0}{ |a_0| }} e^{\nu} \right), \qquad a_0<0 \,, \\
\rm \Psi &=& \rm \psi_0 e^{\frac{1}{\hbar}\left(\delta_1 \xi + \delta_2 \eta + \delta_3 \chi + \delta_4 \rho -\frac{b_1}{2a_0}\nu\right)} J_\mu\left(
\frac{2}{\hbar}\sqrt{\frac{2V_0}{ a_0 }} e^{\nu} \right), \qquad a_0>0 \,, 
\end{eqnarray}
where $\rm K_\mu$ and $\rm J_\mu$ are the modified and the ordinary Bessel functions, respectively. On the other hand, for the case $\rm a_{0}=0$, the equation to solve is 
\begin{equation}
\rm\frac{dG}{d\nu} + \left(\frac{8V_0}{b_1 \hbar} e^{2\nu}- \frac{b_2}{b_1 \hbar} \right)G=0 \,,
\end{equation}
whose solution becomes
\begin{equation}
\rm G=G_0 e^{ \frac{b_2}{b_1 \hbar}\nu} Exp\left[-\frac{4V_0}{b_1\hbar}e^{2\nu}\right]\,,  \end{equation}
and thus the wave function yields
\begin{equation}
\rm \Psi=\psi_0 e^{\frac{1}{\hbar}\left(\delta_1 \xi + \delta_2 \eta + \delta_3 \chi +\delta_4 \rho +\frac{b_2}{b_1}\nu\right)}Exp\left[-\frac{4V_0}{b_1\hbar}e^{2\nu}\right] \,,\quad a_{0}=0 \,.   
\end{equation}
In any case the wave function exhibits a damping behaviour with respect to the variable $\rm\nu$, which represents the combination of the scale factors $\rm\{a,b,c\}$ and the fields $\rm\{\phi,\sigma\}$. Indeed, this is an anticipated feature.

\section{Conclusions\label{conclusions}}

We analyzed the case of Bianchi type I model with multi-scalar fields cosmology. We introduced the corresponding EKG system of equations and the associated Hamiltonian density. Exact solutions to the EKG system are derived by means of Hamilton's approach where a particular scalar potential of the form $\rm V=  V_0 e^{-\left[\lambda_1 \phi_1 + \cdots + \lambda_n \phi_n \right]}$ was utilized, which gave rise to different cases dependant of the free  parameter $\rm \lambda$, for which the scalar fields, the scale factors and the e-folding function were found. In particular, we found a simple solution for the case $\rm\lambda^2=3$, where 60 e-folds can be achieved. The Hamiltonian density was employed in order to compute the WDW equation, which was solved by means of a change of variables and an adequate ansatz; yielding three distinct solution depending of particular choices of the $\rm \lambda$ parameter, moreover, all of them exhibit a damping behaviour with respect to the scale factors and the fields, which is a sought aspect of the quantum scheme.     

Even though the anisotropic cosmological models, represent an enticing prospect to explain the early stages of the universe, and no conclusive evidence has been found that a primordial anisotropy is needed. However, they still remain an interesting topic of study, given that if a signal in the thermal maps of the CMB is found to be connected to an anisotropic spacetime, then such models, which are more featured rich, would reveal a deeper understanding of the hidden/unknown aspects of the early universe, even those predating inflation. 

Even though, few attempts have been made to scrutinize the observable imprints of anisotropic spacetime, and while no conclusive evidence of such stamp has been found yet, nor the contrary neither; even more for the multi-field scenario. However, the aim of this paper lies outside of such thriving ideas. Wistfully they still remain as a pending study.

\acknowledgments{
\noindent \noindent This work was partial supported by PROMEP grant
UGTO-CA-3. This work is part of the collaboration within the
Instituto Avanzado de Cosmolog\'{\i}a. Many calculations where done
by Symbolic Program REDUCE 3.8. RHJ and NEO acknowledges CONACyT for financial support.}

\end{document}